\documentclass[aps,twocolumn,showpacs]{revtex4}
\usepackage{graphicx}
\usepackage{epsfig}
\usepackage{epstopdf}
\usepackage{amsfonts}
\usepackage{amssymb}
\usepackage{amsbsy}
\usepackage{amsmath}
\usepackage{mathrsfs}
\usepackage{latexsym}
\usepackage{natbib}
\usepackage{bm}
\usepackage{color}
\usepackage{braket}
\usepackage{slashed}
\usepackage{pgfplots}
\usepackage[normalem]{ulem}

\usepackage{tikz}
\usetikzlibrary{shapes,arrows,shadows}
\def\x{\mathbf{x}}

\def\k{\mathrm{k}}

\def\mut{\tilde{\mu}}
\def\mt{\bar{m}}

\def\muimb{\bar{\mu}_{Im}}

\def\bt{\bar{b}}

\def\rhot{\tilde{\rho}}

\def\mr{\mathcal{M}}
\def\mtl{\tilde{M}}

\begin{document}

\title{Equation of states in the curved spacetime of spherical degenerate  
stars}

\author{Golam Mortuza Hossain}
\email{ghossain@iiserkol.ac.in}

\author{Susobhan Mandal}
\email{sm17rs045@iiserkol.ac.in}

\affiliation{ Department of Physical Sciences, 
Indian Institute of Science Education and Research Kolkata,
Mohanpur - 741 246, WB, India }
 

\date{\today}

\begin{abstract}

In the study of spherical degenerate stars such as neutron stars, general 
relativistic effects are incorporated by using Tolman-Oppenheimer-Volkoff 
equations to describe their interior spacetime. However, the equation of states 
employed in such studies are invariably computed in flat spacetime. We show that 
the equation of states computed in the curved spacetime of these stars depend 
explicitly on the metric function. Further, we show that ignoring such 
metric-dependent gravitational time dilation effect leads one to grossly 
underestimate the mass limits of these compact stars.

\end{abstract}

\maketitle

\emph{Introduction.--}\label{Introduction}
The observation of gravitational waves, accompanied by electromagnetic 
counterparts \cite{TheLIGOScientific:2017qsa,GBM:2017lvd}, has opened up an 
unprecedented window to probe several uncharted aspects of matter 
field dynamics in  a strong gravity regime. These observed gravitational waves 
are thought to have originated from the merger of neutron stars which are 
supported by fermionic degeneracy pressure. The general relativity plays a key 
role for these stars due to their compact size and significant mass. In order 
to incorporate the effects of general relativity in their mass-radius 
relations, one usually employs the Tolman-Oppenheimer-Volkoff (TOV) equations 
to describe the interior spacetime. However, the equation of states (EOS) 
used in such studies, are invariably computed using the flat spacetime (see 
 \cite{Lattimer:2012nd,Ozel:2016oaf,Baym:2017whm} for recent reviews).

The usage of flat metric locally can be justified while computing the EOS 
at a given radial location within a star. However, two such locally 
inertial frames located at two different radial locations are \emph{not} 
identical as their clock speeds differ due to the \emph{gravitational time 
dilation}. Therefore, an EOS computed in the flat spacetime (flat EOS) cannot 
capture the effects of strong gravity on the matter field dynamics within these 
stars. In turns, it necessitates a first principle derivation of the EOS using 
the curved spacetime (curved EOS) of these stars.

\emph{Fermions in curved spacetime.--}
In order to keep the analysis simple yet fairly general, here we consider the 
non-interacting MIT Bag model \cite{Chodos:1974je} to describe the degenerate 
matter within a star. Inside the bag region and ignoring the boundary terms, 
the matter field action in a curved spacetime with the metric $g_{\mu\nu}$ can 
be expressed as $S_{Bag} = S_{QCD} - \int d^4x \sqrt{-g} B$ where $B$ is the Bag 
constant. By relying on the asymptotic freedom, the QCD action is further 
approximated as $S_{QCD} = \sum_{I} S_{\psi_I}$ where the index $I$ runs over 
different types of 4-component Dirac spinor field $\psi_I$. We may emphasize 
here that by setting $B=0$ and choosing the spinor field $\psi_I$ to represent 
either the electrons or the neutrons, the analysis here would directly describe 
the degenerate matter within an ideal white dwarf or a neutron star 
respectively.

In Minkowski spacetime, the fermions are described by the spinor representation 
of the Lorentz group. In order to describe the fermions in the curved spacetime,
here we follow the Fock-Weyl formulation where the Dirac action is invariant 
under both the general coordinate transformations as well as the \emph{local} 
Lorentz transformations. At every point in a curved spacetime, a set of local 
coordinates, say $\xi^{a}$, can be defined in which the metric in the global 
coordinates $x^{\mu}$, say $g_{\mu\nu}$, becomes Minkowski metric as 
$g_{\mu\nu} {e^{\mu}}_a {e^{\nu}}_b = \eta_{ab}$ with $\eta_{ab} = 
diag(-1,1,1,1)$ being the Minkowski metric. We use Greek letters to denote the 
indices of global coordinates whereas Latin letters are used for locally  
inertial coordinates. The \emph{tetrad} components ${e^{\mu}}_a \equiv 
({\partial x^{\mu}}/{\partial \xi^{a}})$ which relate the global frame and the 
locally inertial frame, transform under the general coordinate transformation, 
$x^{\mu}\rightarrow x'^{\mu}$, as a \emph{contra-vector} and under the local 
Lorentz transformation, $\xi^{a}\rightarrow \xi'^{a}$, as a \emph{co-vector}. 
Using the tetrad and the \emph{inverse} tetrad ${e_{\mu}}^{a}$, one can relate 
the components of any contra-variant vector field $V^{\mu}$ or co-variant 
vector field $V_{\mu}$, between the global frame and the local frame as 
follows
\begin{equation}\label{TetradCoContraVector}
\tilde{V}^{a} = {e_{\mu}}^{a} V^{\mu} ~~,~~
\tilde{V}_{a} = {e^{\mu}}_{a} V_{\mu} ~.
\end{equation}
Under local Lorentz transformation ${\Lambda^a}_b(x) \equiv 
(\partial\xi'^a/\partial\xi^b)$, a fermion field $\psi(x)$ transform in the 
spinor representation as $\psi (x)\rightarrow \psi'(x) = U[\Lambda(x)]\psi(x)$. 
In the curved spacetime $\partial_{\mu}U[\Lambda(x)] \ne 0$ in general. 
Consequently, the term $\partial_{a}\psi$ does not transform as a co-vector 
under the local Lorentz transformation \emph{i.e.} $\partial_{a}\psi$ 
$\rightarrow$ $\partial_{a}'\psi' \ne {({\Lambda^{-1})}_{a}}^b U[\Lambda] 
\partial_{b} \psi$.
Therefore, in order to extend the Dirac action in the curved spacetime, a 
suitable covariant derivative for the spinor field is defined as 
$\tilde{\mathcal{D}}_{a} \psi \equiv {e^{\mu}}_{a} \mathcal{D}_{\mu}\psi \equiv 
{e^{\mu}}_{a} [\partial_{\mu}\psi + \Gamma_{\mu}\psi]$ such that under local 
Lorentz transformation it leads to
\begin{equation}\label{CovariantDerivative}
\tilde{\mathcal{D}}_{a}\psi \rightarrow {({\Lambda^{-1})}_{a}}^b 
U[\Lambda] ~\tilde{\mathcal{D}}_{b}\psi  ~.
\end{equation}
In order to derive the form of $\Gamma_{\mu}$, we consider an infinitesimal 
local Lorentz transformation given by ${\Lambda^a}_b = \delta^a_b + 
{\Omega^a}_b$ where $\Omega_{ab} =- \Omega_{ba}$. In spinor representation, 
$U[\Lambda] = \mathbb{I} + \frac{1}{2} \Omega_{ab} \sigma^{ab}$ where 
$\sigma^{ab}=\frac{1}{4}[\gamma^{a},\gamma^{b}]$ with  $\gamma^{a}$ being the 
Dirac matrices in the locally inertial frame. The Dirac matrices $\gamma^{a}$ 
satisfy $\{\gamma^{a},\gamma^{b}\} = - 2\eta^{ab} \mathbb{I}$. The minus sign in 
front of $\eta^{ab}$ is chosen here so that for given metric signature, the 
Dirac matrices satisfy the usual relations $(\gamma^0)^2 = \mathbb{I}$ and 
$(\gamma^k)^2 = -\mathbb{I}$ where $k=1,2,3$. 
The action of the covariant derivative on the Dirac adjoint $\bar{\psi} = 
\psi^{\dagger} \gamma^0$ is given by $\mathcal{D}_{\mu}\bar{\psi} \equiv  
[\partial_{\mu}\bar{\psi} - \bar{\psi}\Gamma_{\mu}]$ which ensures 
$\bar{\psi}\psi$ behaves as a scalar under both local Lorentz transformation as 
well as general coordinate transformation. Further, one demands that the 
covariant derivative $\mathcal{D}_{\mu}$ must act on 
$(\bar{\psi}\gamma^{\nu}\psi)\equiv (\bar{\psi}\gamma^{a} {e^{\nu}}_a\psi)$ 
which is a contra-vector under general coordinate transformation, as the regular 
covariant derivative $\nabla_{\mu}$ \emph{i.e.} 
$\mathcal{D}_{\mu}(\bar{\psi}\gamma^{\nu}\psi) = 
\partial_{\mu}(\bar{\psi}\gamma^{\nu}\psi) + \Gamma^{\nu}_{\mu\beta} 
(\bar{\psi}\gamma^{\beta}\psi)$ where $\Gamma^{\nu}_{\mu\beta}$ is the 
Christoffel connection. 
Now it is convenient to express $\Gamma_{\mu} \equiv -\tfrac{1}{2} 
\omega_{\mu ab} \sigma^{ab}$. Subsequently, the use of Leibniz rule for 
both covariant derivative and partial derivative together with compatibility 
conditions for the tetrad \emph{i.e.} $\mathcal{D}_{\mu}  
{e^{\nu}}_{a} = 0 = \mathcal{D}_{\mu} {e_{\nu}}^{a}$ leads to 
\begin{equation}\label{OmegaMuabDef}
\omega_{\mu ab}  = \eta_{ac} {e_{\nu}}^c \left[ \partial_{\mu} {e^{\nu}}_b 
+ \Gamma^{\nu}_{\mu\sigma} {e^{\sigma}}_b \right] ~.
\end{equation}
We note that $\gamma^{\mu} \equiv \gamma^{a} {e^{\mu}}_a$ satisfy 
$\{\gamma^{\mu},\gamma^{\nu}\} = - 2g^{\mu\nu} \mathbb{I}$ and are often 
referred to as the Dirac matrices in the curved spacetime.
Therefore, in the Fock-Weyl formulation, the invariant 
action for the $I^{th}$ spinor in the curved spacetime which is minimally 
coupled with the geometry, can be expressed as
\begin{equation}\label{FermionActionI}
S_{\psi_I} = -\int d^{4}x \sqrt{-g} ~ \bar{\psi}_I [i 
\gamma^{a} {e^{\mu}}_a \mathcal{D}_{\mu} + m_I]\psi_I  ~,
\end{equation}
where $\bar{\psi}_I = \psi^{\dagger}_I \gamma^0$ and $m_I$ is its mass. The 
field equation for the $I^{th}$ Dirac spinor in the curved spacetime is given by
\begin{equation}\label{FermionEOMI}
[i \gamma^{a} {e^{\mu}}_a \mathcal{D}_{\mu} + m_I]\psi_I = 0 ~.
\end{equation}
The corresponding conservation equation is $\mathcal{D}_{\mu} j^{\mu}_I = 0$ 
where 4-current density is given by $j^{\mu}_I = \bar{\psi}_I\gamma^{a} 
{e^{\mu}}_a \psi_I$.

\emph{Stress-energy tensor and equation of state.--}
The key idea behind application of quantum field theory in the curved 
background is to consider the Einstein equation of the form 
$G_{\mu\nu} = 8\pi G \langle \hat{T}_{\mu\nu}\rangle$. Here the Einstein 
tensor $G_{\mu\nu}$ is treated classically whereas the classical stress-energy 
tensor $T_{\mu\nu}$ is replaced by the appropriate expectation value of the 
corresponding quantum operator $\hat{T}_{\mu\nu}$. Assuming \emph{perfect fluid} 
form for the stress-energy tensor, here it would imply 
$\langle\hat{T}_{\mu\nu}\rangle = (\rho+P)u_{\mu}u_{\mu} + P g_{\mu\nu}$.
On the other hand, the stress-energy tensor corresponding to a spinor field 
$\psi$ is given by
\begin{equation}\label{TMuNuPsiDef}
T_{\mu\nu} = - \frac{ {e_{(\mu a}}}{\sqrt{-g}}
\frac{\delta S_{\psi}}{\delta {e^{\nu)}}_{a} } ~,
\end{equation}
which reduces to the standard form $T_{\mu\nu} = - \frac{2}{\sqrt{-g}} 
\frac{\delta S_{\psi}}{\delta g^{\mu\nu}}$. Using on-shell condition, the Eq. 
(\ref{TMuNuPsiDef}) leads to ${T}^0_0 = - \mathcal{H}$ and ${T}^k_k = 
\mathcal{H} - m \bar{\psi}\psi$ where the field Hamiltonian is $H = \int d^{3}x 
\sqrt{-g} \mathcal{H}$. The corresponding partition function in a given small 
region is
\begin{equation}\label{PartitionFunctionTr}
\mathcal{Z}_{\psi} = \text{Tr}[e^{-\beta(\hat{H} - \mu\hat{N})}]  ~,
\end{equation}
where $\mu$ is chemical potential, $\hat{N}$ is number operator, $\beta = 1/k_B 
T$ with $T$ and $k_B$ being the temperature and the Boltzmann constant 
respectively. Using the Eq. (\ref{PartitionFunctionTr}), one arrives at the 
usual form of the energy density 
$\rho = \langle\hat{H}\rangle/V$ with volume of the small region $V = \int 
d^{3}x\sqrt{-g}$, and $\langle\hat{H} - \mu\hat{N}\rangle = - (\partial \ln 
\mathcal{Z}_{\psi}/\partial\beta)$ where $\langle\cdot\rangle$ denotes 
the thermal expectation value. Further, $\ln\mathcal{Z}_{\psi}$ being a 
dimensionless \emph{extensive} quantity, it can be expressed in the form
$\ln\mathcal{Z}_{\psi} = \beta^{-3} V f(\beta\mu,\beta m)$ (for eg. Eq. 
(\ref{LogPartitionFunctionI})) where 
$\beta\frac{\partial}{\partial\beta}f(\beta\mu,\beta m) =
m \frac{\partial}{\partial m}f(\beta\mu,\beta m) 
+ \mu\frac{\partial}{\partial \mu}f(\beta\mu,\beta m)$. Consequently, 
the corresponding pressure takes the standard form $P = (\beta V)^{-1} 
\ln\mathcal{Z}_{\psi}$.
We may emphasize here that the partition function (\ref{PartitionFunctionTr}) 
is \emph{generally} invariant for the given \emph{external} parameters
$\beta$ and $\mu$. However, what is often overlooked in the literature that 
these parameters which set the scale of energy density and pressure, are 
impacted by the general relativistic \emph{time dilation} effects due to 
radially varying lapse function within a star. In this article, we show the 
results of time dilation effect on the equation of state by two methods: direct 
computation and using the scaling behaviour of the time coordinate.

\emph{Spacetime within spherical stars.--}
Using \emph{natural units}, $c = \hbar =1$, the invariant line element within a 
spherical star can be written as
\begin{equation}\label{InteriorMetric}
ds^2 = - e^{2\Phi(r)}dt^2 + e^{2\nu(r)} dr^2 + 
r^2(d\theta^2 + \sin^2\theta d\phi^2) ~.
\end{equation}
The metric functions $\Phi(r)$ and $\nu(r)$ are determined by solving 
the Einstein equations, also referred to as the TOV equations. In particular, 
these equations lead to $e^{-2\nu(r)} = (1 - 2 G \mr/r)$ and
\begin{equation}\label{TOVEqn}
\frac{d\Phi}{dr} = \frac{G(\mr + 4\pi r^3 P)}{r(r - 2 G \mr)} ~~,~~
\frac{dP}{dr} = - (\rho + P) \frac{d\Phi}{dr} ~,
\end{equation}
where $d\mr = 4\pi r^2\rho dr$. 
Inside the star, the pressure $P$ and the energy density $\rho$ both vary 
radially. On the other hand, at thermodynamic equilibrium, the quantities such 
as the pressure, the energy density are uniform within the system. In order to 
combine these two aspects together, one considers a sufficiently small spatial 
region, containing sufficient degrees of freedom, around a given point 
within the star where \emph{local} thermodynamic equilibrium holds.

\emph{Reduced spinor action.--}
It is always possible to find a coordinate system in which the 
metric is locally flat and which can be used to describe the nuclear 
interactions. Here we give such an explicit construction where we also 
retain the information about the metric function $\Phi$.
For definiteness, let us a consider a small box whose center is located at a 
radial coordinate $r_0$. One may expand the metric functions $\Phi(r)$ and 
$\nu(r)$ around the point $r_0$ and keep only the leading terms. Further, given 
the spherical symmetry, we may rotate the coordinate system such that the polar 
axis, $\theta=0$, passes through the center of the box. Then for all points 
within the box, the angle $\theta$ can be taken to be \emph{small}. By using a 
new set of coordinates
$X = e^{\nu(r_0)} r \sin \bar{\theta} \cos\phi$, 
$Y = e^{\nu(r_0)} r \sin \bar{\theta} \sin\phi$, and
$Z = e^{\nu(r_0)} r \cos \bar{\theta}$ along with $\bar{\theta} =  
e^{-\nu(r_0)}\theta$, the metric within the box can be reduced to
\begin{equation}\label{MetricInTOVBox}
ds^2 = - e^{2\Phi(r_0)}dt^2 + dX^2 + dY^2 + dZ^2 ~.
\end{equation}
The Eq. (\ref{MetricInTOVBox}) shows that the metric is flat  within the box, 
\emph{i.e.} it is flat over a scale which is sufficient to describe the nuclear 
interactions. At the same time, it shows that the metric is not \emph{globally} 
flat as it carries information about the large scale radial variation of the 
metric function $\Phi$, implied by the TOV Eqs. (\ref{TOVEqn}). This is in 
contrast to the usage of globally flat metric for computation of EOS in the 
literature. By using the diagonal ansatz, the corresponding tetrad and inverse 
tetrad can be expressed as
\begin{equation}\label{TetradInTOVBox}
{e_{\mu}}^{a} = diag(e^{\Phi},1,1,1)  ~,~
{e^{\mu}}_{a} = diag(e^{-\Phi},1,1,1)  ~,
\end{equation}
where $\Phi \equiv \Phi(r_0)$. Clearly, in the $(t,X,Y,Z)$ coordinates both 
$\Gamma^{\mu}_{\alpha\beta}$ and $\omega_{\mu ab}$ vanish and the 
spin-covariant derivative becomes $\mathcal{D}_{\mu} = \partial_{\mu}$ within 
the box. The action (\ref{FermionActionI}) for the spinor $\psi_I$ then reduces 
to
\begin{equation}\label{FermionActionReduced}
S_{\psi_I} = -\int d^{4}x ~\bar{\psi}_I \left[ i \gamma^{0} \partial_0 
+ e^{\Phi} \left(i \gamma^{k} \partial_{k} + m_I\right) \right]\psi_I  ~,
\end{equation}
where $k$ runs over $1,2,3$. The corresponding conserved charge then becomes 
$Q_I = \int d^3x \sqrt{-g} j^0_I = \int d^3x \bar{\psi}_I \gamma^{0} \psi_I$.
The reduced action (\ref{FermionActionReduced}) should be viewed as an 
\emph{effective} field action in a locally Minkowski spacetime. In contrast to 
the standard spinor action used in the literature, the action 
(\ref{FermionActionReduced}) carries information about the box-specific,  
\emph{fixed} metric function $\Phi$ and hereafter we use this action for 
computation of the equation of state.

\emph{Partition function.--}
In order to compute the EOS, here we employ the tools of thermal quantum field 
theory \cite{matsubara1955new,kapusta2006finite} as pioneered by Matsubara. The 
partition function corresponding to the MIT Bag model that represents the spinor 
degrees of freedom within the box is given by
\begin{equation}\label{LogPartitionFunctionBag}
\ln\mathcal{Z}_{Bag} = \sum_I \ln\mathcal{Z}_{\psi_I}  - e^{\Phi} \beta V B ~,
\end{equation}
where $V=\int d^{3}x$ denotes the 
spatial volume of the box. In path-integral formulation, the 
partition function for the $I^{th}$ spinor is $\mathcal{Z}_{\psi_I} = 
\int \mathcal{D}\bar{\psi}_I \mathcal{D}\psi_I e^{-S_{\psi_I}^{\beta}}$ where 
its Euclideanized action is
\begin{equation}\label{PartitionFunctionSBE}
S_{\psi_I}^{\beta} = \int_0^{\beta}d\tau \int d^3\x~ 
\left[\mathcal{L}_{\psi_I}^{E} + \mu_I \bar{\psi}_I(\tau,\x) 
\gamma^0\psi_I(\tau,\x) \right] ~,
\end{equation}
with $\mu_I$ being its chemical potential. The Euclideanized Lagrangian 
density $\mathcal{L}_{\psi_I}^{E}$ which is obtained through the substitution 
$t\to i\tau$ from its standard counterpart, is
\begin{equation}\label{EuclideanLagrangianDensity}
\mathcal{L}_{\psi_I}^{E} = - \bar{\psi}_I(\tau,\x)[ \gamma^{0} \partial_{\tau} 
+ e^{\Phi} \left(i \gamma^{k} \partial_{k} + m_I\right) ]\psi_I(\tau,\x)  ~.
\end{equation}
The \emph{anti-periodic} boundary condition which carries the information about 
the equilibrium temperature $T$, is imposed on the spinor field as 
\begin{equation}\label{FermionicBoundaryCondition}
\psi_I(\tau,\x) = -\psi_I(\tau+\beta,\x)  ~.
\end{equation}
It is convenient to transform the spinor field in the Fourier domain as
\begin{equation}\label{FermionicFourier}
\psi_I(\tau,\x) = \frac{1}{\sqrt{V}} \sum_{n,\k} ~e^{i(\omega_n\tau + 
\k\cdot\x)} \tilde{\psi}(n,\k)  ~.
\end{equation}
The anti-periodic boundary condition (\ref{FermionicBoundaryCondition}) leads 
the Matsubara frequencies to be $\omega_n = (2n+1) \pi ~\beta^{-1}$ where 
$n$ is an integer. The Eqs. (\ref{PartitionFunctionSBE}) and 
(\ref{FermionicFourier}) together lead to
\begin{equation}\label{PartitionFunctionSBEFourier}
S_{\psi_I}^{\beta} = \sum_{n,\k} ~\bar{\tilde{\psi}}~\beta
\left[ \slashed{p} - \mt_I \right]  \tilde{\psi} ~,
\end{equation}
where $\mt_I =  m_I e^{\Phi}$,  $\slashed{p} = \gamma^{0}(-i\omega_n+\mu_I) + 
\gamma^{k} (\k_k e^{\Phi})$
and the corresponding thermal propagator in the Fourier domain is given 
by
\begin{equation}\label{SpinorPropagator}
\mathcal{G}_{I}(\omega_n,\k) = \frac{1}{\slashed{p} - \mt_{I}} ~.
\end{equation}
Using the Dirac representation of the gamma 
matrices and the result of Gaussian integral over the Grassmann variables one 
gets
\begin{eqnarray}\label{LogPartitionFunctionFermion2}
\ln\mathcal{Z}_{\psi_I} = 2 \sum_{\k} \ln \left(1 + e^{-\beta(\omega-\mu_I)} 
\right) ~,
\end{eqnarray}
where $\omega = \omega(\k) = e^{\Phi} \sqrt{\k^2+m_I^2}$. Here here we have 
ignored formally divergent zero-point energy and the anti-particle 
contributions.

A degenerate star is characterized by the condition $\beta\mu_I \gg 1$. It 
allows one to approximate $({e^{\beta(\omega - \mu_I)} + 1})^{-1} \simeq 
\Theta(\mu_I-\omega) - \mathrm{sgn}(\mu_I-\omega) e^{-\beta|\mu_I-\omega|}$ 
where $\Theta(x)$ and $\mathrm{sgn}(x)$ are Theta and signum functions 
respectively. Using this approximation, one can carry out the summation 
over $\k$ label by converting it to an integral to get
\begin{equation}\label{LogPartitionFunctionI}
\ln\mathcal{Z}_{\psi_I} = \frac{\beta V e^{-3\Phi}}{24\pi^2} \left[ 
2\mu_I \mu_{Im}^3 - 3\mt_I^2 \muimb^2 
+ \frac{48 \mu_I \mu_{Im}}{\beta^{2}} \right] ~,
\end{equation}
where  $\mu_{Im} \equiv \sqrt{\mu_I^2-\mt_I^2}$ and $\muimb^2 \equiv 
\mu_I\mu_{Im} - \mt_I^2 \ln(\tfrac{\mu_I + \mu_{Im}}{\mt_I})$. In the limit 
$\Phi\to 0$, $\ln\mathcal{Z}_{\psi_I}$ reduces to the standard flat spacetime 
form (see \cite{Hossain:2019teg,Hossain:2019qpx}).

\emph{Equation of state.--}
The number density of the $I^{th}$ spinor can be computed using $n_I = (\beta 
V)^{-1} (\partial \ln \mathcal{Z}_{Bag} /\partial \mu_I)$ as
\begin{equation}\label{ElectronNumberDensityFull}
 n_I = \frac{\mu_{Im}^3 e^{-3\Phi}}{3\pi^2} \left[1 + 
\frac{6}{(\beta\mu_{Im})^2} \left(1 + \frac{\mu_I^2}{\mu_{Im}^2} \right)
\right] ~,
\end{equation}
where we have used the relations $(\partial \muimb^2/\partial\mu_I) = 
2\mu_{Im}$ and $(\partial\mu_{Im}/\partial\mu_I) = (\mu_I/\mu_{Im})$.
For a grand canonical ensemble, total pressure $P = (\beta V)^{-1}\ln 
\mathcal{Z}_{Bag}$ leads to
\begin{equation}\label{PressureFullBag}
P =  \sum_I \frac{e^{-3\Phi}}{24\pi^2} \left[ 
2\mu_I \mu_{Im}^3 - 3\mt_I^2 \muimb^2 + \frac{48 \mu_I \mu_{Im}}{\beta^{2}}
 \right] -e^{\Phi} B ~.
\end{equation}
We may parameterized the baryon number density as $n = (n_I/\bt_I)$ with $\bt_I$ 
being the model-dependent parameters. Given $\beta\mu_I \gg 1$ for a degenerate 
star, the temperature corrections of $\mathcal{O}((\beta\mu_I)^{-2})$ are very 
small and henceforth ignored for simplicity. Then the Eq. 
(\ref{ElectronNumberDensityFull}) leads to
\begin{equation}\label{DensityRelations}
\mu_{Im} e^{-\Phi} = m_I (b_I n)^{1/3} ~,~
\mu_I e^{-\Phi} = m_I \sqrt{(b_I n)^{2/3} + 1} ~,~
\end{equation}
where the constant $b_I = 3\pi^2 \bt_I/m_I^3$. The pressure $P$ can be 
expressed in terms of baryon number density $n$ as
\begin{eqnarray}\label{BagPressureCurved}
P &=& e^{\Phi}\sum_I\frac{m_I^4}{24\pi^2} \left[ \sqrt{(b_I n)^{2/3} + 1}
\left\{2(b_I n) - 3(b_I n)^{1/3} \right\} \nonumber \right. \\
&+& \left. ~3 \ln\left\{ (b_I n)^{1/3}+\sqrt{(b_I n)^{2/3} + 
1} \right\} \right]  - e^{\Phi} B ~. 
\end{eqnarray}
The energy density $\rho$ within the box can be computed using $(\rho - \sum_I 
\mu_I n_I)V = -(\partial \ln \mathcal{Z}_{Bag} /\partial \beta)$ and
ignoring temperature corrections as earlier, it leads to
\begin{equation}\label{BagEnergyDensityCurved}
\rho = -P + e^{\Phi} \sum_I \frac{m_I^4}{3\pi^2} \sqrt{(b_I n)^{2/3} + 1}
~ (b_I n) ~.
\end{equation}
The key property of the curved EOS (\ref{BagPressureCurved}, 
\ref{BagEnergyDensityCurved}) is its explicit dependence on $\Phi$ which 
follows from the metric component $g_{tt}$. Clearly, it is the 
\emph{gravitational time dilation} effect which is experienced by the matter 
field in a strong gravitational field and is missed in a derivation using flat 
spacetime. We may emphasize that the EOS is computed here in a small box located 
around $r=r_0$. However, $r_0$ being arbitrary, the EOS can be considered to be 
dependent on $r$. 

\emph{Numerical solutions.--} 
In order to solve the TOV Eqs. numerically, we note that both pressure and 
energy density are explicit functions of $\Phi$ as $P = P(n,\Phi)$ and $\rho = 
\rho(n,\Phi)$. This in turns implies that TOV Eqs. (\ref{TOVEqn}) can be viewed 
as a set of first order differential equations for the triplet $\{\mr,\Phi,n\}$ 
where the baryon density $n$ satisfies
\begin{equation}\label{BaryonNumberEqn}
\frac{dn}{dr} = - \frac{(\rho + P + ({\partial P}/{\partial \Phi}))}
{({\partial P}/{\partial n})} \frac{d\Phi}{dr} ~.
\end{equation}
We note that for the curved EOS, unlike for the flat EOS, the metric function 
$\Phi$ can not be eliminated from the set of TOV Eqs. (\ref{TOVEqn}). 
Nevertheless, for a star of radius $R$ and mass $M$, the triplet 
$\{\mr,\Phi,n\}$ is subject to the conditions $e^{2\Phi(R)} = (1 - 2G M /R)$ 
with $M = \mr(R)$ and $n(R) = 0$. In order to numerically impose this condition, 
for a given central baryon density, say $n_c$, we evolve these equations towards 
the surface with a trial value of $\Phi$ at the center. We independently 
calculate the value $\Phi$ at the surface, say $\Phi_s = 
\tfrac{1}{2}\ln(1 - 2G M /R)$ and compare it with the evolved value of $\Phi$ 
at the surface. In the next step, we compute $\Phi$ at the center starting from 
the computed value $\Phi_s$ and by evolving it backward from $n=0$ to $n=n_c$ 
using the Eq. 
\begin{equation}\label{dndPhiEqn}
\frac{d\Phi}{dn} = - \frac{({\partial P}/{\partial n})} {(\rho + P + 
({\partial P}/{\partial \Phi}))} ~,
\end{equation}
which follows from the Eq. (\ref{BaryonNumberEqn}). These steps are 
then iterated until the evolved and computed values of the metric function 
$\Phi$ at the surface agree with each other within the desired numerical 
accuracy. The above method of iteration leads to a rapid convergence on the 
value of $\Phi$.
\emph{Speed of sound and causality.--}
We note that the pressure and energy density for the curved EOS can be 
written as $P = P(n,\Phi)$ and $\rho = \rho(n,\Phi)$ whereas
for the flat EOS they are of the form $\tilde{P} = \tilde{P}(n)$ and 
$\tilde{\rho} = \tilde{\rho}(n)$. By defining the respective speed of sound as
$c_s^2 = dP/d\rho$ and $\tilde{c}_s^2 = d\tilde{P}/d\tilde{\rho}$ together with 
the Eq. (\ref{dndPhiEqn}), one can establish a relation as follows
\begin{equation}\label{SpeedOfSound}
c_s^2 = \tilde{c}_s^2 \frac{\tilde{\rho}+\tilde{P}}
{(1-\tilde{c}_s^2)\tilde{\rho} + 2\tilde{P}}  ~.
\end{equation}
In the high baryon density limit \emph{i.e.} $n\to\infty$ both $c_s^2$ and
$\tilde{c}_s^2$ approaches $\tfrac{1}{3}$ from below but maintaining 
$c_s^2 > \tilde{c}_s^2$ for $0 < n < \infty$. In other words, for a given 
baryon density speed of sound for the curved EOS is higher compared to the flat 
EOS but it remains well below the speed of light.

\emph{Enhanced mass limits.--}
In contrast to the energy density in flat spacetime, say $\rhot = \rhot(n)$, 
the energy density (\ref{BagEnergyDensityCurved}) is of the form $\rho = 
\rho(n,\Phi)$ where $\tfrac{d \Phi}{dr} > 0$ within a star except at the 
center. It implies that $\rho(r + \Delta r) > \rhot(r + \Delta r)$ even if 
$\rho(r) = \rhot(r)$ and $(\tfrac{\partial \rho}{\partial n} \tfrac{d 
n}{dr})_{|r} = (\tfrac{\partial \rhot}{\partial n} \tfrac{d n}{dr})_{|r}$ for 
positive $\Delta r$. Secondly, the TOV Eqs. (\ref{TOVEqn}) imply that a higher 
central energy density leads to a faster fall-off in pressure which in turns 
leads to a smaller radius of the star. Therefore, if one aims to have two stars 
of same radius, one using curved EOS and another using flat EOS, then the 
central energy density for the curved EOS must be higher.

Let us consider two sets of solutions of the TOV equations, one using curved EOS 
denoted as $\{\mr^{C},\Phi^{C},n^{C}\}$ and another using flat EOS denoted as
$\{\mr^{F},\Phi^{F},n^{F}\}$ such that $n^{C}(R) = 0 = n^{F}(R)$ and $\rho(R) = 
0 = \rhot(R)$. The latter condition does not permit a non-zero $B$ to be present 
in lower densities. It then follows from the arguments of the previous 
paragraph that energy densities satisfy $\rho(n^{C},\Phi^{C}) > \rhot(n^{F})$ 
for $r<R$. The masses corresponding to the curved and flat EOS, say $M$ and 
$\mtl$ respectively, then satisfy $M > \mtl$. For baryon number density $n\to 
0$, the Eqs. (\ref{TOVEqn}) implies $\tfrac{d n}{dr} \sim - n^{1/3} 
\tfrac{d\Phi}{dr}$ for both the cases whereas the energy densities vary as $\rho 
\sim n^{C} e^{\Phi^{C}}$ and $\rhot \sim n^{F}$. Therefore, in the limit 
$r\to R$, $\rho > \rhot$ implies $(\tfrac{d\Phi^{C}}{dr} e^{2\Phi^{C}/3} )_{|R} 
> (\tfrac{d\Phi^{F}}{dr})_{|R}$ which together with the TOV Eqs. (\ref{TOVEqn}) 
leads to 
\begin{equation}\label{MassRelation}
M > \mtl \left[1 + \frac{2G\mtl}{3R}- \frac{80 (G\mtl)^3}{81 R^3} + 
\mathcal{O}\left(\tfrac{G^4\mtl^4}{R^4}\right) \right] ~.
\end{equation}
The mass inequality (\ref{MassRelation}) implies that the usage of flat EOS 
leads one to \emph{underestimate} the masses of the degenerate stars (see the 
figures for quantitative comparison).

\begin{figure}
\centering
\includegraphics[width=8cm]{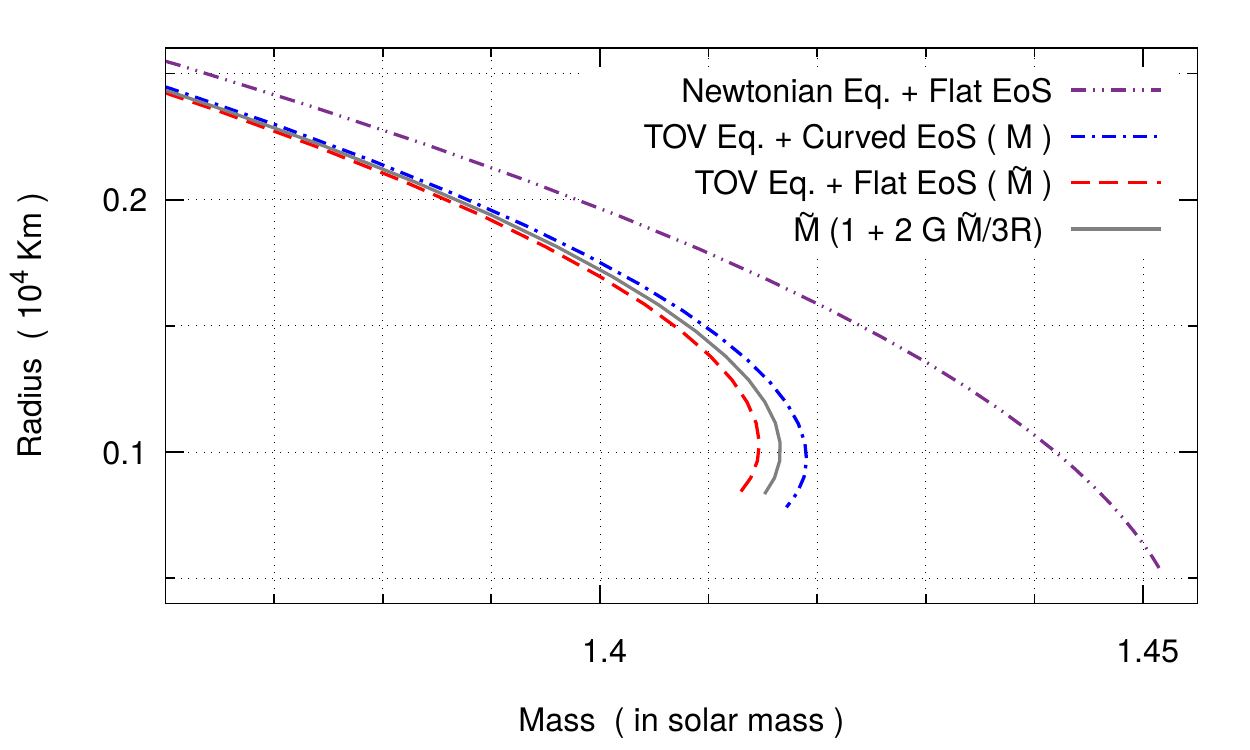}
\caption{The mass-radius relations for the white dwarfs with $A/Z=2$. The 
curved EOS leads the maximum mass limit to increase from around $1.415 
M_{\odot}$ to $1.419 M_{\odot}$}.
\label{fig:mass-radius-comparison-wd}
\end{figure}

\emph{White dwarfs.--}
The EOS for a white dwarf can be read off from Eq. (\ref{BagPressureCurved}) by 
choosing the fermions to be the electrons, and by setting $B=0$, $\bt_I = 
Z/A$ where $A$, $Z$ are the atomic mass number and atomic number respectively. 
In the ultra-relativistic limit, such an EOS reduces to the standard polytropic 
form, up to the factor of $e^{\Phi}$, as
\begin{equation}\label{PressureWDUV}
P \simeq \frac{e^{\Phi}}{12\pi^2} \left(\frac{3\pi^2 Z ~n}{A}  
\right)^{4/3}  ~.
\end{equation}
In white dwarfs, the nuclei have negligible motion. For such a nucleus within 
the given box, the invariant action $S_n = - m_u A\int\sqrt{-ds^2}$ reduces to 
$S_n = \int dt[-m_u A e^{\Phi}]$ where $m_u$ is the atomic mass unit. The 
corresponding Hamiltonian then implies that the energy density due to the nuclei 
is $n m_u e^{\Phi}$ which needs to be included in total energy density 
$\rho$. The numerically evaluated mass-radius relations for the white dwarfs are 
plotted in the FIG. \ref{fig:mass-radius-comparison-wd}. 

\begin{figure}
\centering
\includegraphics[width=8cm]{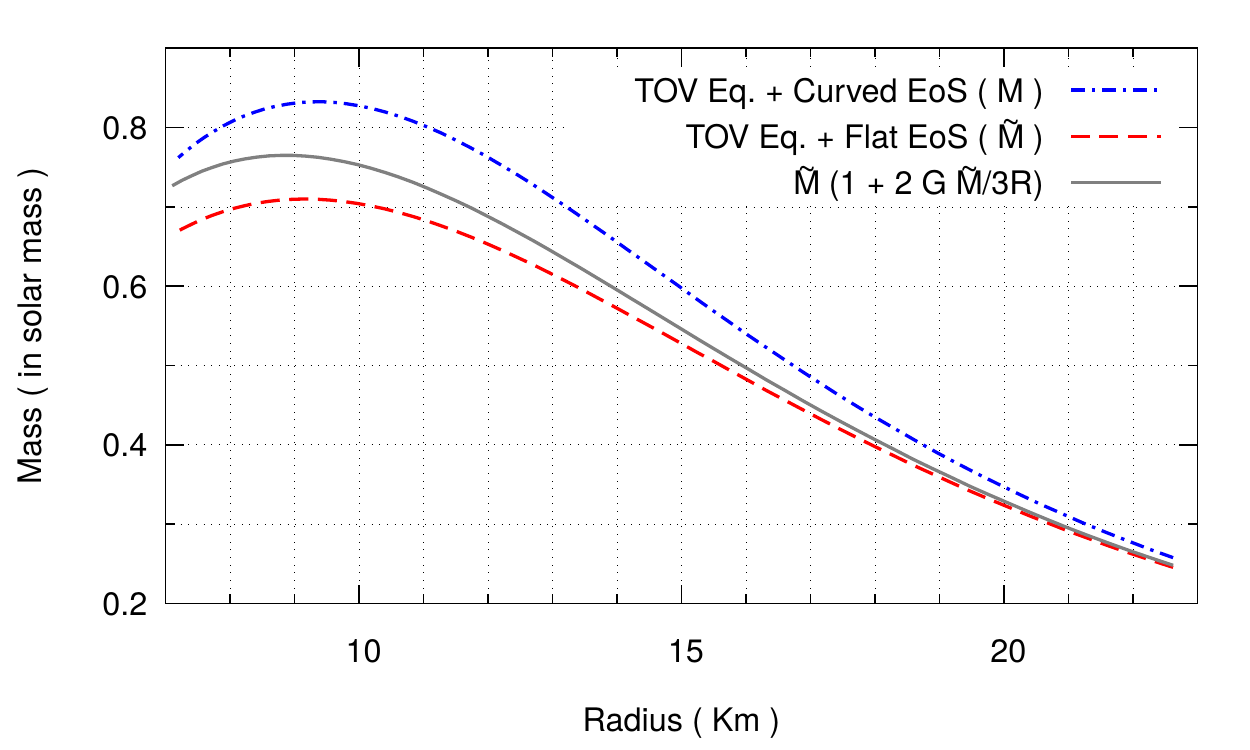}
\caption{The mass-radius relations for ideal neutron stars. The maximum mass 
limit increases by $\sim 16.9\%$, from around $0.71 M_{\odot}$ to $0.83 
M_{\odot}$ and the corresponding radius increases by $\sim 2.2\%$, from 
approximately $9.2$ km to $9.4$ km, due to the usage of curved EOS. }
\label{fig:mass-radius-comparison-ns}
\end{figure}

\emph{Neutron stars.--}
The EOS for an \emph{ideal} neutron star follows from the Eqs. 
(\ref{BagPressureCurved}, \ref{BagEnergyDensityCurved}), if one chooses the 
fermions to be the neutrons, and sets $B=0$, $\bt_I = 1$. The inclusion of 
gravitational time dilation effect on the EOS, leads the mass limit of an ideal 
neutron star to increase by $\sim 16.9 \%$ which is significantly higher than 
the minimum increase ($\sim 7.5 \%$) implied by the Eq. (\ref{MassRelation}) 
(see FIG. \ref{fig:mass-radius-comparison-ns}).

For high density neutron stars, there are strong stability arguments suggesting 
that the nuclear matter should consist of hyperons \emph{i.e.} baryons 
containing \emph{strange} quark. Unfortunately, many such models stumble 
to explain the observed high mass ($\gtrsim 2 M_{\odot}$) neutron stars 
\cite{Demorest:2010bx,Antoniadis:2013pzd,Cromartie:2019kug}. This conundrum 
is referred to as the \emph{hyperon puzzle} \cite{Bombaci:2016xzl}.
Nevertheless, there exist other proposed models that include hyperons and can 
lead to observed high mass of the neutron stars \cite{zdunik2013maximum, 
maslov2015solution, bednarek2012hyperons}. In any case, the Eq. 
(\ref{MassRelation}) implies that ignoring the effects of 
the curved spacetime on the EOS leads one to grossly underestimate the mass 
limits of these compact stars. For example, if a flat EOS leads the mass 
limit to be, say, $1.5 M_{\odot}$ with a radius of $10$ Km then the 
corresponding curved EOS would enhance the mass limit by $\sim 13.7\%$ at the 
minimum. While the actual increase of the mass limit can be obtained only 
through a comprehensive computation, it is expected to be comparatively higher 
as in the case of ideal neutron star. In any case, the legitimate 
incorporation of the curved spacetime while deriving neutron star EOS would 
significantly alleviate the hyperon puzzle even for those models which lead to 
relatively lower mass limits. Furthermore, the usage of the curved EOS would 
imply even higher mass limits for those hyperon models that otherwise can 
explain the observed high masses of the neutron stars.

In the quark matter models, the nuclear matter is made up of \emph{up}, 
\emph{down} and \emph{strange} quarks each having 3 \emph{color} degrees of 
freedom. Having more species of fermions for a given baryon density, quark 
matters lead to higher pressure and the higher mass limits. At equilibrium, 
the masses of the up and down quarks are negligible \emph{i.e.} $m_u 
\ll \mu_u$, $m_d \ll \mu_d$ whereas strange quark mass is small
\emph{i.e.}  $m_s < \mu_s$. By considering the index $I$ to run over these 
different types of quarks, we can express the total pressure 
(\ref{PressureFullBag}) as $P \simeq  e^{\Phi} \left[e^{-4\Phi}\left(\mu_u^4 + 
\mu_d^4 + \mu_s^4 - 3 \mu_s^2 \mt_s^2 \right)/4\pi^2 -B \right]$. Based on the 
equilibrium interactions, one can fix the relative strength of different 
chemical potentials. The \emph{strange star} scenario \cite{alcock1986strange} 
is obtained by choosing $\mu \equiv \mu_u = \mu_d \approx \mu_s$ along with 
$\bt_I=1$, as
\begin{equation}\label{PressureExpandedStrangeRho}
P \simeq  \frac{3 e^{\Phi}}{4\pi^2} 
\left\{ a_4 (\mu e^{-\Phi})^4 - a_2 (\mu e^{-\Phi})^2 \right\} -B e^{\Phi} ~,
\end{equation}
where the parameters $a_4=1$ and $a_2 = m_s^2$. The so-called \emph{color-flavor 
locked} (CFL) phase scenario \cite{rajagopal2001enforced} requires that these 
quarks have same number density implying their respective chemical potentials to 
satisfy $\mu_u = \mu_d = \sqrt{\mu_s^2 - \mt_s^2} \equiv \mu$. The corresponding 
pressure again can be expressed in the form (\ref{PressureExpandedStrangeRho}) 
with $a_4=1$ and $a_2 = m_s^2/3$. As the condition $\rho(R)=0$ cannot be 
satisfied with non-zero bag constant $B$, the quark matter EOS must be stitched 
together with the appropriate hadronic EOS in low density as done in the well 
known equations of states such as by Akmal, Pandharipande, and Ravenhall, 
\cite{Akmal:1998cf}, Togashi et al. \cite{Togashi:2017mjp}, Baym et al. 
\cite{Baym:2019iky}. Nevertheless, even the quark matter EOS 
(\ref{PressureExpandedStrangeRho}) depends explicitly on the metric function 
$\Phi$. We may mention that inclusion quark \emph{interactions}, perturbatively, 
would simply alter the values of $a_2$ and $a_4$.

\emph{Curved EOS from flat EOS.--}
Owing to uncertain theoretical understanding of the nuclear matter, numerous 
neutron star EOS have been studied in the literature. Invariably, these EOS are 
computed in the flat spacetime. We now show that it is possible to convert a 
flat EOS into its spherically symmetric curved spacetime counterpart. Firstly, 
within the box  one may define a new time coordinate $\tilde{t} = e^{\Phi} t$ 
which leads the metric (\ref{MetricInTOVBox}) to become the standard Minkowski 
metric $ds^2 = - d\tilde{t}^2 + dX^2 + dY^2 + dZ^2$. Let us denote the 
equilibrium temperature and chemical potential, as seen from the frame 
$(\tilde{t},X,Y,Z)$, to be $\tilde{T} = 1/(k_B \tilde{\beta})$ and $\mut_I$ 
respectively. As seen from the frame $(t,X,Y,Z)$, the information about $T$ is 
contained within the anti-periodic boundary condition 
(\ref{FermionicBoundaryCondition}) with period $\Delta\tau = \beta$. Now time 
intervals of these two frames are related as $\Delta\tilde{t} = e^{\Phi} \Delta 
t$ which implies
\begin{equation}\label{BetaRelation}
\tilde{\beta} = e^{\Phi} \beta ~.
\end{equation}
The chemical potential satisfies $(k_B \beta)\mu_I = - (\partial S/ 
\partial N_I)$ where $S$ is the entropy and $N_I$ is the particle number of the 
$I^{th}$ spinor in the box. At equilibrium, the entropy $S$ and particle 
number $N_I$ are not affected by a scaling of the time coordinate which implies
\begin{equation}\label{MuRelation}
\mut_I = e^{-\Phi} \mu_I  ~.
\end{equation}
One can verify using the relations ($\ref{BetaRelation}$, $\ref{MuRelation}$) 
that the partition function evaluated using standard Minkowski metric, becomes 
the same as in the Eq. (\ref{LogPartitionFunctionBag}).

\emph{Discussions.--}
To summarize, the clock speeds of two locally inertial frames located at 
different radial coordinates within a spherical star, differ due to the 
\emph{gravitational time dilation} effect. The metric (\ref{MetricInTOVBox}) 
which is used here, explicitly shows that although one can use a flat metric to 
compute an EOS locally, \emph{i.e.} at the scale of nuclear interactions, the 
metric itself is not globally flat. In contrast, in the existing literature, the 
EOS is computed in a globally flat metric and is used for solving TOV Eqs 
(\ref{TOVEqn}) for all values of radial coordinates. It leads one to overlook 
the effects of time dilation on the EOS caused by the large scale radial 
variation of the metric function $\Phi$. We have shown that ignoring such 
metric-dependent time dilation effect on the EOS of neutron stars, leads one to 
grossly underestimate their mass limits. Given the flat EOS are widely used in 
the neutron star literature, the result presented here would imply significant 
alterations of various existing predictions.

\emph{Acknowledgments:} SM would like to thank IISER Kolkata for supporting this 
work through a doctoral fellowship. 

% 
% \bibliographystyle{apsrev}
% \bibliography{bibtexfile}
%

\end{document}